\documentclass[amsmath,amssymb,preprint,showpacs]{revtex4}

\usepackage{graphicx}
\usepackage[usenames]{color}

\begin{document}

\title{Charge dynamics of the spin-density-wave state in BaFe$_2$As$_2$}
\author{F. Pfuner$^1$, J.G. Analytis$^2$, J.-H. Chu$^2$, I.R. Fisher$^2$ and L. Degiorgi$^1$} \affiliation{$^1$Laboratorium f\"ur
Festk\"orperphysik, ETH - Z\"urich, CH-8093 Z\"urich,
Switzerland}
\affiliation{$^2$Geballe Laboratory for Advanced Materials and
Department of Applied Physics, Stanford University, Stanford,
California 94305-4045, U.S.A.}

\date{\today}

\begin{abstract}
We report on a thorough optical investigation of BaFe$_2$As$_2$ over a broad spectral range and as a function of temperature, focusing our attention on its spin-density-wave (SDW) phase transition at $T_{SDW}=135$ K. While BaFe$_2$As$_2$ remains metallic at all temperatures, we observe a depletion in the far infrared energy interval of the optical conductivity below $T_{SDW}$, ascribed to the formation of a pseudogap-like feature in the excitation spectrum. This is accompanied by the narrowing of the Drude term consistent with the $dc$ transport results and suggestive of suppression of scattering channels in the SDW state. About 20\% of the spectral weight in the far infrared energy interval is affected by the SDW phase transition.

\end{abstract}

\pacs{75.30.Fv,78.20.-e}


\maketitle

A new field in condensed matter research was recently initiated by the discovery of superconductivity in doped iron oxyarsenides \cite{Kamihara}. After the first report on LaFeAs(O$_{1-x}$F$_x$) with a critical temperature $T_c$ of 26 K, even higher transition temperatures up to 55 K were reached in fluorine doped SmFeAs(O$_{1-x}$F$_x$) \cite{Kamihara,Chen}. This novel family of compounds has generated a lot of interest, primarily because high-temperature superconductivity is possible in materials without CuO$_2$ planes. Since these planes are essential for superconductivity in the copper oxides, its occurrence in the pnictides raises the tantalizing question of a different pairing mechanism. While the nature of such a pairing mechanism is currently in dispute, it is evident that superconductivity in the oxypnictides emerges from specific structural and electronic conditions in the (FeAs)$^{\delta-}$ layer. Therefore, it soon appears that other structure types could serve as parent material. The recently discovered BaFe$_2$As$_2$ compound with the well-known ThCr$_2$Si$_2$-type structure is an excellent candidate \cite{Rotter}. This latter structure contains practically identical layers of edge-sharing FeAs$_{4/4}$ tetrahedra with the same band filling, but separated by barium atoms instead of LaO sheets. BaFe$_2$As$_2$ can serve as an alternative compound for oxygen-free iron arsenide superconductors. Superconductivity was indeed established at 38 K in K-doped \cite{Rotter2} and at 22 K in Co-doped \cite{Sefat} BaFe$_2$As$_2$ through partial substitution of the Ba  and Fe site, respectively.

Superconductivity in suitably doped BaFe$_2$As$_2$ compounds conclusively proves that it originates only from the iron arsenide layers, regardless of the separating sheets. It is also currently believed that the superconductivity in these systems is intimately connected with magnetic fluctuations and a spin-density-wave (SDW) anomaly within the FeAs layers. This is obviously of interest, since a SDW phase may generally compete with other possible orderings and complicated phase diagrams are often drawn due to their interplay. Undoped LaFeAsO as well as BaFe$_2$As$_2$ undergo a SDW phase transition which is also associated with a reduction of the lattice symmetry from tetragonal to orthorhombic \cite{Kamihara,Rotter}. The classical prerequisite for the formation of a SDW state is the (nearly) perfect nesting of the Fermi surface (FS), leading to the opening of a gap. Density functional calculations of the electronic structure \cite{Singh} reveal that the main effect of doping is to change the relative sizes of the electron and hole FS and therefore to lead to a reduction in the degree of nesting of the FS itself. Doping suppresses the SDW instability and induces the superconductivity. Alternatively, it was also proposed that the metallic SDW state could be solely induced by interactions of local magnetic moments (i.e., without invoking the opening of a gap), resembling the nature of antiferromagnetic order in the cuprate parent compounds \cite{Yang}. It is therefore of relevance to acquire deeper insight into the SDW phase and to establish how this broken symmetry ground state affects the electronic properties of these materials.

We focus here our attention on BaFe$_2$As$_2$, which is a poor metal exhibiting Pauli paramagnetism and where the SDW phase transition at $T_{SDW}$ is accompanied by anomalies in the specific heat, electrical resistance and magnetic susceptibility \cite{Rotter}. We provide a thorough investigation of the optical response over an extremely broad spectral range at temperatures both above and below $T_{SDW}$. A self-consistent picture emerges in which below the SDW transition a reshuffle of states \cite{band} occurs due to the opening of a SDW gap on a portion of the FS (i.e., pseudogap). We also observe a mobility increase of the charge carriers, which are not gapped by the SDW condensate, as a consequence of the reduction of electron-electron scattering in the SDW state.


BaFe$_2$As$_2$ single crystals were grown by a self flux method as described in Refs. \onlinecite{xstal1} and \onlinecite{xstal2}. Ba and FeAs in the molar ration 1:4 were placed in an alumina crucible and
sealed in quartz tube. The mixture was heated to 1190 $^{\circ}$C, and held for
32 hours, before it was slowly cooled to 1070 $^{\circ}$C, at which temperature the
furnace was powered off. After cooling to room temperature
single crystals were mechanically separated from the solidified ingot. The large ($\sim$ 2x2 mm$^2$)
crystals have a platelet morphology, with the $c$-axis perpendicular
to the plane of the platelet. Our specimens were characterized through $dc$ transport investigations (see inset Fig. 3, below), obtaining equivalent results as reported in Ref. \onlinecite{Rotter} and indicating a SDW transition at $T_{SDW}\sim$ 135 K. We measured the optical reflectivity $R(\omega)$ of shiny surfaces from the far infrared (FIR) up to the ultraviolet at temperatures from 300 K down to 10 K. Our specimen was placed inside an Oxford cryostat with appropriate optical windows. From the FIR up to the mid-infrared spectral range we made use of a Fourier-transformed interferometer equipped with a He-cooled bolometer detector, while in the visible and ultraviolet energy intervals we employed a Perkin Elmer spectrometer. Measuring $R(\omega)$ over such a broad spectral range allows to perform reliable Kramers-Kronig (KK) transformation from where we obtain the phase of the complex reflectance and consequently all optical functions. Details pertaining to the experiment and the KK analysis can be found elsewhere \cite{Wooten,Gruner-Dressel}.

\begin{figure}[!tb]
\center
\includegraphics[width=7.5cm]{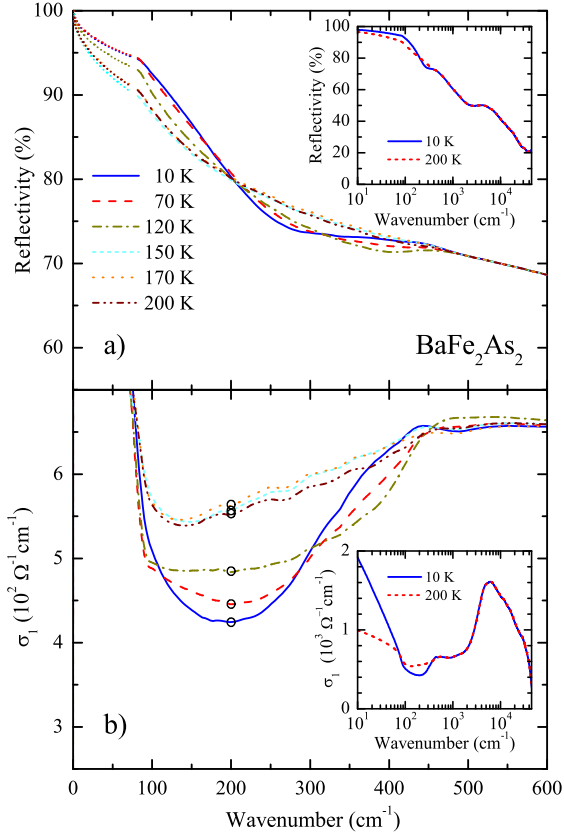}
\caption{(color online) (a) Optical reflectivity and (b) real part $\sigma_1(\omega)$ of the optical conductivity as a function of temperature in the FIR spectral range. The dotted lines in panel (a) emphasize the Hagen-Rubens extrapolation of $R(\omega)$ towards zero frequency. The insets in both panels show the same quantities at 10 and 200 K from the far infrared up to the ultraviolet spectral range with logarithmic energy scale. The open circles in panel (b) at 200 cm$^{-1}$ mark the values of $\sigma_1$ used for the calculation of the $\Lambda$ quantity (see text).} \label{Refl.}
\end{figure}

Figure 1a displays the optical reflectivity in the FIR energy interval (i.e., $\omega\le$ 600 cm$^{-1}$), emphasizing its temperature dependence. The inset of Fig. 1a shows on the other hand $R(\omega)$ over the whole measured range at 10 and 200 K. $R(\omega)$ of BaFe$_2$As$_2$ has an overall metallic behavior, characterized by a broad bump at about 5000 cm$^{-1}$ and by the onset of the reflectivity plasma edge below 3000 cm$^{-1}$ (inset Fig. 1a). Of interest is the temperature dependence of $R(\omega)$, which first indicates a depletion of $R(\omega)$ with decreasing temperature in the energy interval between 200 and 500 cm$^{-1}$ and then a progressive enhancement of $R(\omega)$ below 200 cm$^{-1}$ and towards zero frequency, so that all $R(\omega)$ spectra cross around 200 cm$^{-1}$. Apart from the already mentioned broad excitation at 5000 cm$^{-1}$, we did not observe any additional mode in $R(\omega)$ below 3000 cm$^{-1}$ and overlapped to its plasma edge. This contrasts with previous optical data \cite{Dong,Tropeano} of the SDW phase of La(O$_{1-x}$F$_x$)FeAs and SmFeAs(O$_{1-x}$F$_x$), where several sharp modes in the infrared range were indeed detected. These features were ascribed to the vibrational lattice modes of both the $ab$ plane as well as of the orthogonal $c$ axis. The presence of both polarizations is in agreement with the polycrystalline nature of those samples \cite{Dong,Tropeano}, suggesting also that the main contribution to the optical properties of pellets comes from the more insulating $c$ axis. The absence of (sharp) phonon peaks in our spectra indicates that we are addressing the conducting ($ab$) plane, where the phonon modes are screened by the effective Drude component. In this regard, our $R(\omega)$ spectra share common features with the optical findings on K-doped BaFe$_2$As$_2$ \cite{Li} and LaFePO \cite{Basov} single crystals in the normal phase. Furthermore, it is worth mentioning that we did not find any polarization dependence of $R(\omega)$ within the platelets surface. Therefore, this reinforces the notion that we are addressing the isotropic ($ab$) plane of BaFe$_2$As$_2$. 

For the purpose of the KK transformation, the $R(\omega)$ spectra may be extended towards zero frequency with the well-known Hagen-Rubens extrapolation \cite{Wooten,Gruner-Dressel} $R(\omega$)=$1-2\sqrt{\omega/\sigma_{dc}}$ (Fig. 1a), making use of $\sigma_{dc}$ values in fair agreement with the transport result (inset Fig. 3). At high frequencies $R(\omega)$ can be extrapolated as $R(\omega)\sim\omega^{-s}$ (2 $< s <$ 4) \cite{Wooten, Gruner-Dressel}. Both types of extrapolations, besides being standard extension of the measured data, do not affect either the discussion of or the conclusions drawn from our findings. Figure 1b then shows the real part $\sigma_1(\omega)$ of the optical conductivity at various temperatures, resulting from the KK transformation of the measured $R(\omega)$. In the inset of Fig. 1b we display the overall optical conductivity at 10 and 200 K.

The first interesting feature in $\sigma_1(\omega)$ is the peak centered at about 5000 cm$^{-1}$, with a high frequency shoulder extending between 10$^4$ to 3x10$^4$ cm$^{-1}$ (inset Fig. 1b). We ascribe that wealth of excitations to the electronic interband transitions involving Fe $d$-states and As $p$-states \cite{Singh}. The low frequency tail of this peak merges at high temperatures into a rather broad interval of excitations extending in the mid-infrared range of $\sigma_1(\omega)$. The main focus of our discussion is however on the two most prominent behaviors, developing below 600 cm$^{-1}$ with decreasing temperature, as emphasized in the main panel of Fig. 1b. There is first an obvious depletion in $\sigma_1(\omega)$, which is reminiscent of a (pseudo)gap opening, and second the narrowing of the Drude component below 150 cm$^{-1}$. The onset of that depletion as well as the Drude narrowing clearly occur at temperatures below $T_{SDW}$. These latter features characterizing $\sigma_1(\omega)$ in BaFe$_2$As$_2$ are common to the other oxypnictides \cite{Dong,Tropeano,Basov} as well as to the electrodynamic response of high temperature superconducting cuprates \cite{BasovHTC}.

We apply a phenomenological Lorentz-Drude fit in order to better single out the various components, contributing to $\sigma_1(\omega)$ and shaping the electrodynamic response of the title compound, as well as to disentangle the distribution of spectral weight among them \cite{Wooten,Gruner-Dressel}. The FIR spectral range is recovered with a Drude term, accounting for the effective metallic component, and three temperature dependent Lorentz harmonic oscillators (h.o.), describing the absorptions at finite frequency. Figure 2a highlights these phenomenological contributions to $\sigma_1(\omega)$ and particularly emphasizes the spectral weight of the three h.o.'s, covering the FIR energy interval. Additional h.o.'s (not shown), which turn out to be temperature independent, are also considered in order to reproduce the high-frequency part of the excitation spectrum. Figure 2b displays the spectral weight of the three FIR h.o.'s ($S_i^2$, the square of the h.o. strength) as well as of the Drude term ($\omega_p^2$, the square of the plasma frequency). The Drude weight remains basically constant or at most weakly decreases between 200 and 10 K. On the other hand, the strongest redistribution of weight occurs between the h.o. component at about 170 cm$^{-1}$, and the two adjacent h.o.'s. There is indeed a transfer of weight from the range between 100 and 300 cm$^{-1}$ into the high frequency tail of the Drude term, represented by the h.o. at about 50 cm$^{-1}$, and the absorption around 420 cm$^{-1}$ (Fig. 2a). The total spectral weight, encountered in the Drude term and the three FIR h.o.'s of Fig. 2a, remains however constant at all temperatures (Fig. 2b) \cite{comment}.

\begin{figure}[!tb]
\center
\includegraphics[width=7.5cm]{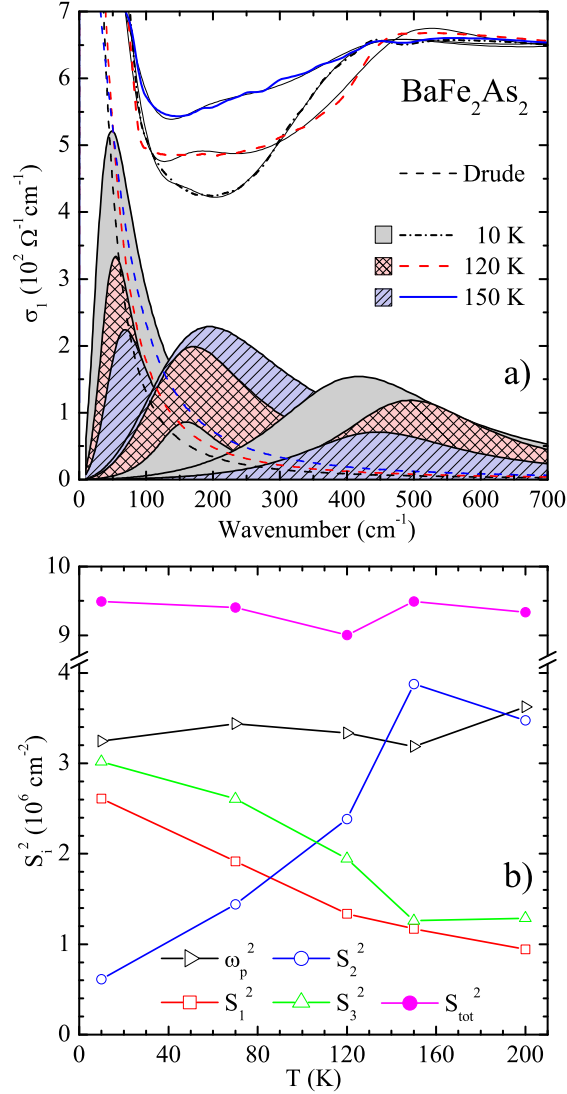}
\caption{(color online) (a) Measured $\sigma_1(\omega)$ from Fig. 1b and Lorentz-Drude components at three selected temperatures in the FIR. For each temperature the dashed line is the Drude term while the shaded areas emphasize the spectral weight of the FIR h.o.'s, centered at $\omega_1=$ 49 cm$^{-1}$, $\omega_2=$ 160 cm$^{-1}$ and $\omega_3=$ 417 cm$^{-1}$ at 10 K. The thin lines show the total fit. (b) Temperature dependence of the spectral weights for the Drude component ($\omega_p^2$) and for the Lorentz h.o.'s ($S_i^2$) in the FIR energy interval as well as of the total weight ($S_{tot}^2=\omega_p^2+\sum_{i=1}^3
S_i^2$).} \label{spectral}
\end{figure}

Such a redistribution of spectral weight governs the appearance of the depletion in $\sigma_1(\omega)$ between 100 and 400 cm$^{-1}$ below $T_{SDW}$ (Fig. 1b). Following a well established procedure \cite{Sacchettiprb}, we thus define the characteristic average weighted energy scale:
\begin{equation}
\omega_{SP}=\frac{\sum_{j=2}^3 \omega_j S_j^2}{\sum_{j=2}^3
S_j^2},
\end{equation}
where $\omega_2$ and $\omega_3$ are the resonance frequencies of the second and third h.o. in our Lorentz-Drude fit (Fig. 2a). The quantity $\omega_{SP}$ is then ascribed to the representative energy scale for the pseudogap excitation, resulting below $T_{SDW}$ from the reshuffle of states at energies lower than the mobility edge. The ratio between $\omega_{SP}$ at 10 K and $k_BT_{SDW}$ ($k_B$ being the Boltzmann constant) is of about 4, somehow larger than the prediction within the weak-coupling limit of the mean-field BCS theory \cite{BCS} but comparable to what has been found in the SDW organic Bechgaard salts \cite{Vescoli}. $\omega_{SP}$ is shown in Fig. 3, normalized by its value at 10 K. The increase of $\omega_{SP}$ below $T_{SDW}$ is obvious. This goes hand in hand with the decrease of $\sigma_1(\omega)$ in the energy interval around 200 cm$^{-1}$ (open circles in Fig. 1b). We calculate the quantity $\Lambda=\sigma_1$(200 cm$^{-1}$,T)/$\sigma_1$(200 cm$^{-1}$,10 K), and find indeed that $\Lambda^{-1}$ scales with $\omega_{SP}$ as a function of temperature (Fig. 3).

As far as the Drude term is concerned, its narrowing (Fig. 1b) is well represented by the decrease of the scattering rate parameter ($\Gamma_D$) with decreasing temperature. The inset of Fig. 3 displays $\Gamma_D$ from our Lorentz-Drude fit (Fig. 2a) as a function of temperature together with $dc$ resistivity data taken for samples prepared by the same technique. Both quantities are normalized to their respective values at 200 K and follow the same trend in temperature above as well as below $T_{SDW}$. The SDW transition reduces the possible scattering channels so that the resistivity and the Drude scattering rate decrease rather abruptly below $T_{SDW}$. 

 \begin{figure}[!tb]
\center
\includegraphics[width=7.5cm]{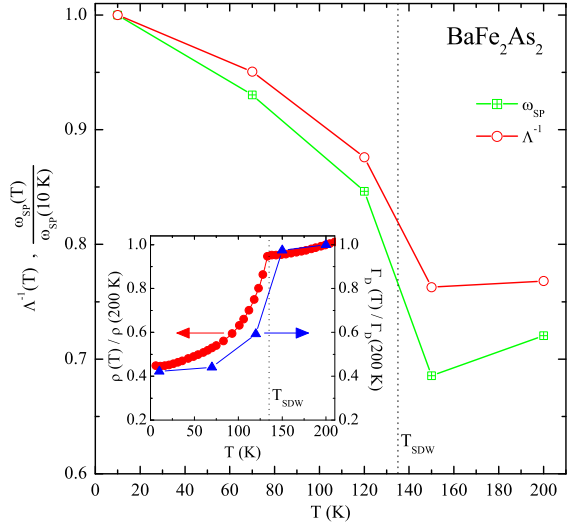}
\caption{(color online) Temperature dependence of the pseudogap energy scale $\omega_{SP}$ normalized at 10 K and of the quantity $\Lambda^{-1}$ (see text). The inset shows the temperature dependence of the $dc$ resistivity and of the Drude scattering rate $\Gamma_D$ normalized by their values at 200 K.} \label{gap.}
\end{figure}

Therefore, from the data as well as from their phenomenological Lorentz-Drude analysis emerges a picture in which the metallic component of the electrodynamic response is not just a simple Drude term. At $T>T_{SDW}$ delocalized states (i.e., free charge carriers close to the Fermi level), represented by the Drude term, coexist with more localized or low-mobility states, described in $\sigma_1(\omega)$ by the three FIR Lorentz h.o.'s covering the spectral range from 50 up to 700 cm$^{-1}$ (Fig. 2a). When lowering the temperature below $T_{SDW}$, a pseudogap develops and a rearrangement of states takes place in such a way that excitations pile up in the spectral range around 420 cm$^{-1}$ as well as around 50 cm$^{-1}$. These latter excitations merge into the high frequency tail of the narrow Drude term of $\sigma_1(\omega)$. 
Comparing the removed spectral weight within the depletion of $\sigma_1(\omega)$ in FIR below $T_{SDW}$ with the total spectral weight encountered in $\sigma_1(\omega)$ up to about 700 cm$^{-1}$ above $T_{SDW}$, we estimate that approximately 20\% of the total weight is affected by the development of the pseudogap.

In conclusion, we have provided a comprehensive analysis of the excitation spectrum in BaFe$_2$As$_2$, with respect to its SDW phase transition. Our data reveal the formation of a pseudogap excitation and the narrowing of the metallic contribution in $\sigma_1(\omega)$ at $T_{SDW}$, which tracks the behavior of the $dc$ transport properties. It remains to be seen how one can reconcile the pseudogap-like feature well established in the optical response with the apparent absence of any gap excitation in the angle-resolved photoemission spectroscopy \cite{Yang}. This issue is intimately connected to the more general and still open question whether the reshuffle of states, leading to an optical signature compatible with the pseudogap excitation, is the consequence of the SDW-driven structural phase transition and a FS nesting or is in fact due to a more exotic mechanism.\\




The authors wish to thank M. Lavagnini for fruitful discussions. This work has been supported
by the Swiss National Foundation for the Scientific Research
within the NCCR MaNEP pool. This work is also supported by the
Department of Energy, Office of Basic Energy Sciences under
contract DE-AC02-76SF00515.


\begin{thebibliography}{99}
\bibitem{band} BaFe$_2$As$_2$ is a compensated metal and as such we consider all states within a many-band scenario effectively contributing to the transport properties.
\bibitem{comment} The additional weight due to the temperature independent contributions in $\sigma_1(\omega)$ at high frequencies corresponds to an additive constant at all temperatures, which is no longer considered here.
\end{thebibliography}
\end{document}